\documentclass[
    reprint,
    preprintnumbers,
    superscriptaddress,
    nofootinbib,
     amsmath,amssymb,
     aps,
     prd,
    floatfix,
    ]{revtex4-2}

    \usepackage{graphicx}
    \usepackage[usenames,dvipsnames]{xcolor} 
    \usepackage[utf8]{inputenc}
    \usepackage{color}
    \usepackage{hyperref}
    \usepackage[normalem]{ulem} 
    \usepackage{physics}

    \usepackage{enumitem}
    \usepackage{comment}
    \usepackage{bm}
\hypersetup{
  colorlinks   = true, 
  urlcolor     = blue, 
  linkcolor    = blue, 
  citecolor   = blue 
}
    \allowdisplaybreaks[1]
    \graphicspath{{figures/}}

\usepackage{float}

\newcommand{\oxford}{Astrophysics, University of Oxford, DWB, Keble Road, Oxford OX1 3RH, United Kingdom}
\newcommand{\qmul}{Geometry, Analysis and Gravitation, School of Mathematical Sciences, Queen Mary University of London,
Mile End Road, London E1 4NS, United Kingdom}

\newcommand{\phimean}{\langle\phi(r,t)\rangle}
\newcommand{\phimeanbh}{\langle\phi(t)\rangle_\mathrm{bh}}
\newcommand{\phimeanin}{\langle\phi(t)\rangle_\infty}

\begin{document}

\title{Symmetry restoration and vacuum decay from accretion around black holes}

\author{James Marsden}
\email{james.marsden@physics.ox.ac.uk}
\affiliation{\oxford}
\author{Josu C. Aurrekoetxea}
\email{josu.aurrekoetxea@physics.ox.ac.uk}
\affiliation{\oxford}
\author{Katy Clough}
\email{k.clough@qmul.ac.uk}
\affiliation{\qmul}
\author{Pedro G. Ferreira}
\email{pedro.ferreira@physics.ox.ac.uk}
\affiliation{\oxford}

\begin{abstract}
Vacuum decay and symmetry breaking play an important role in the fundamental structure of the matter and the evolution of the universe. In this work we study how the purely classical effect of accretion of fundamental fields onto black holes can lead to shells of symmetry restoration in the midst of a symmetry broken phase. We also show how it can catalyze vacuum decay, forming a bubble that expands asymptotically at the speed of light. These effects offer an alternative, purely classical mechanism to quantum tunnelling for seeding phase transitions in the universe.
\end{abstract}

\maketitle


\noindent
\textbf{\textit{Introduction.}}— Throughout its history, the Universe has undergone a cascade of phase transitions \cite{Peter:2013avv}. 
These transitions have been associated with symmetry breaking, in which the vacuum structure for fundamental fields has morphed from one of higher symmetry to one of lower symmetry \cite{Coleman:1973jx}. It has also been proposed that our universe could, now or in the past, have lived in a false vacuum of some field, and that phase transitions may then be triggered that take it to a true minimum \cite{Coleman:1977py,Callan:1977pt,Coleman:1980aw,Hawking:1981fz}.

The case of quantum transitions between vacuum states has been extensively studied, including the impact of black holes and their potential to enhance the transition rate \cite{Berezin:1987ea,Hiscock:1987hn,Arnold:1989cq,Berezin:1990qs,Gregory:2013hja,Burda:2015yfa,Burda:2015isa,Burda:2016mou,Mukaida:2017bgd,Oshita:2018ptr,Oshita:2019jan}. In this work we will study alternative, purely classical transitions that arise around black holes when they are immersed in coherent scalar fields (such as those that model wave-like dark matter with small mass and high number density, see \cite{Hui:2021tkt} for a review). Such effects are the extreme end of the study of self-interactions in the field, where attractive and repulsive forces arising from the shape of the potential play a role in the classical evolution. In the repulsive case (which we study elsewhere in more detail \cite{Marsden:SIpaper}) the self interactions saturate the growth of density around the black hole (BH), and so the field is confined to one region of the potential. In the attractive case, the growth does not saturate (at least initially), and as a result the field explores greater and greater regions of the potential. It is then affected by the wider shape and in particular the presence of other minima.

The effects that attractive corrections to a quadratic potential have on scalar (and vector) clouds near BHs have previously been considered for the case of self-interacting dark matter like axions \cite{Shapiro:2014oha,Berezhiani:2023vlo,Kadota:2023wlm,Alonso-Alvarez:2024gdz,Boudon:2023vzl,Feng:2021qkj,Alvarez:2020fyo}, including their potential to create so called ``bosenova'' explosions in the case of superradiant clouds \cite{Yoshino:2012kn, Yoshino:2015nsa, Omiya:2022gwu,Omiya:2022mwv, Omiya:2020vji,East:2022rsi,East:2022ppo,Baryakhtar:2020gao,Boskovic:2018lkj} (see \cite{Brito:2015oca} for a review). In the superradiant case, the BH must have a relatively high spin, the scalar mass must satisfy the superradiant condition, and the cloud can dissipate by radiation and by transitions to other bound states from the one being fed by the instability. The accretion case we study is therefore arguably more generic and inescapable, provided a (stable, sufficiently abundant) scalar with some multi-minima potential exists.

Other studies of classical transitions have been related to attempts to model the quantum process as a classical but stochastic process \cite{Linde:1991sk,Brown:2011ry,Braden:2018tky,Blanco-Pillado:2019xny,Hertzberg:2019wgx,Hertzberg:2020tqa,Braden:2022odm,Jenkins:2023npg,Jenkins:2023eez}, with simulations that indicate that the transition (or decay rate) matches the quantum, path integral estimate up to factors of $O(1)$. The process we describe here is inherently classical and deterministic, and simply a consequence of the scalar field evolution on a curved background, rather than an approximation to a quantum system. 

In this work we characterise how the BH accretion process can lead the scalar field to gradually explore multiple vacua. Depending on the form of the potential, this may result in symmetry restoration or trigger a phase transition through vacuum decay. Such processes could be ubiquitous throughout the history of the Universe; BHs exist over a range of mass scales and over a range of times, and may serve as catalysts for such transitions. We consider an isolated Schwarzschild BH, characterising it through numerical simulations and extracting the key features and timescales that impact on observables. 

\vskip 10pt
\noindent
\textbf{\textit{The model.}}— 
In a cosmological setting, phase transitions arise from radiative corrections in field theory, which modify the effective potential seen by the field.\footnote{In the case of a scalar field, Coleman \& Weinberg \cite{Coleman:1973jx} explored how corrections to the potential could modify the vacuum structure, restoring symmetry at high energies. A chemical potential \cite{Schafer:2001bq} or finite temperature corrections \cite{Dolan:1973qd} can have a similar effect.}
In classical field theory we consider the scalar field potential to be fixed and study the dynamics of the field within it. 
We study the dynamics of a real scalar field $\phi$, solving the Klein-Gordon equation
\begin{equation}\label{eq:EKG_cov}
    \nabla^\mu \nabla_\mu \phi - V'(\phi) = 0\,,
\end{equation}
where $\nabla_\mu$ is the covariant derivative operator associated to the metric $g_{\alpha\beta}$. We solve the Klein Gordon equation on a fixed Schwarzschild background, in horizon penetrating Cartesian Kerr–Schild coordinates \cite{Visser:2007fj}
\begin{equation}
   \dd s^2 = \left(\eta_{\mu\nu}+ \frac{2M}{r} l_\mu l_\nu \right) \dd x^\mu \dd x^\nu \,,
\end{equation}
where $\eta_{\mu\nu} =\mathrm{diag}(-1,1,1,1)$ and $M$ is the mass of the black hole.  Here $l_\mu = (1,\, x/r,\, y/r,\, z/r)$ is the ingoing null vector with respect to both $g_{\mu\nu}$ and $\eta_{\mu\nu}$.

To illustrate both the mechanisms of symmetry restoration and vacuum decay, we choose a scalar potential with multiple minima
\begin{equation}\label{eq:scalar_potential}
        V(\phi) = \frac{\lambda}{4}\left(\phi^2-\eta^2\right)^2 + \frac{g}{3}\eta\left(\phi+\eta\right)^3 ~,
\end{equation}
where $\lambda$ and $g$ are dimensionless coupling constants and $\eta$ parameterizes the symmetry breaking scale. The cubic term controls the difference in the energy of each vacuum; in the $g=0$ limit, the potential simplifies to the usual double-well potential with degenerate vacua at $\phi_{\pm} = \pm \eta$. The locations of the minima are given by the parameters of the potential as
\begin{align}
\phi_- &= -\eta\,, \\
\phi_+ &= +\eta\left(\frac{1}{2} + \frac{1}{2}\sqrt{1 - 12x + 4x^2} -x\right)\,,
\end{align}
where $x\equiv (g/2\lambda)$ and the difference in vacuum energy densities is given by $\Delta V\equiv V(\phi_+)-V(\phi_-)$, which is positive if $g>0$ and negative if $g<0$. 

To lowest order around these minima, the potential is  $V(\phi)\approx \mu^2\phi^2/2$, with mass $\mu\approx \sqrt{2\lambda} \eta$. In a flat spacetime, a spatially homogeneous value of the field $\phi$ close to a minimum would oscillate around it with frequency given by the mass $\mu$. Thus, asymptotically far from the black hole the solutions of the scalar field have the form $\phi \propto \cos (\mu t)$.  In the purely massive case, closer to the black hole, the BH gradually accretes a cloud of scalar field \cite{Clough:2019jpm,Bamber:2020bpu}, described by the Heun functions as in \cite{Hui:2019aqm,Vieira:2014waa,Hortacsu:2011rr}, with a power law envelope and characteristic oscillations of the spatial profile on length scales set by the scalar wavelength.
In the regime we consider, the purely massive density profile is proportional to $r^{-3/2}$, with an amplitude that increases over time if our boundary condition continues to feed in more matter. The presence of repulsive self-interactions ($V''>0$), can stabilise the accretion process and prevent the field amplitude from growing \cite{Marsden:SIpaper}. Attractive self-interactions ($V''<0$), on the other hand, result (at least initially, near the minimum of the potential, and as long as there is an asymptotic reservoir of scalar field to feed it) in a continuous growth of the profile, such that closer to the horizon, the amplitude of oscillations increases \cite{Marsden:SIpaper}. This means that the scalar field can explore other features of the potential, such as adjacent vacuum states.

It will be useful to define a mean, time averaged, value of the scalar field at some coordinate radius $r$ 
\begin{equation}
\phimean\approx\frac{1}{T}\int_t^{t+T} \phi(r, t')dt'\,,
\end{equation}
where $T\simeq 2\pi/\mu$ is the local period of the oscillation. If the field oscillates around the negative minimum of the potential $\phimean=-\eta$. 

In this paper we study the dynamics of scalar fields in black hole spacetimes for potentials with multiple minima. We fix $\eta=1$, $\lambda=0.1 M^{-2}$ so that $\mu M \sim 0.5$ and vary $g$ to study symmetry restoration and vacuum decay (we use geometric units $G=c=1$, so that $[\phi] = 1$, $[\rho] = M^{-2}$).  We choose the initial homogeneous value of the scalar field to be $\phi_0=-0.8\eta$, so that it resides near the negative vacuum of a symmetry broken phase. These choices fix the asymptotic energy density to $\rho_0\approx 0.03 \lambda \eta^4$ (for the conversion to physical units see the Discussion).  We perform numerical simulations of the growth of the field using the 3+1 open-source code \textsc{grdzhadzha} \cite{Aurrekoetxea:2023fhl,Andrade:2021rbd}, with octant symmetry to simulate a physical box size of $L=576 M$, a coarsest grid resolution $N^3 = 256^3$ grid points, and 6 levels of refinement, which gives the finest resolution $dx_\mathrm{finest} = 0.035M$ at the BH horizon, located at $r=2M$.

\begin{figure}[t!]
    \centering
    \href{https://youtu.be/c2vcNoX8fVs}{
    \includegraphics[width=\linewidth]{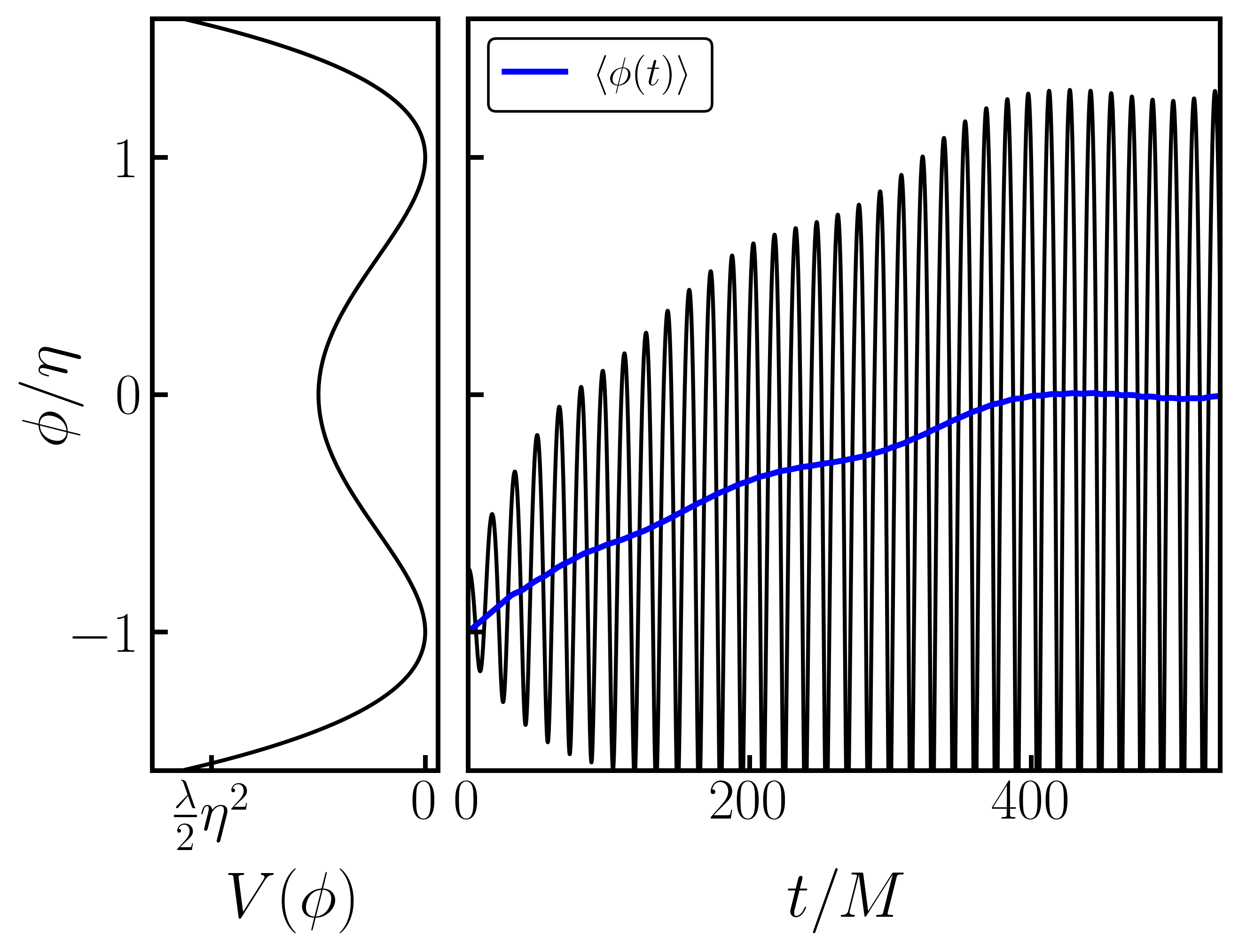}}
    \caption{Symmetry restoration case, where the value of the scalar field at $r=2M$ transitions the barrier and explores the other vacuum (black line). In blue we plot the mean value of the scalar field, which transitions from a symmetry broken phase at $\phimeanbh=-\eta$ to a restored phase at $\phimeanbh=0$.}
    \label{fig:panel_1}
\end{figure}

\vskip 10pt
\noindent
\textbf{\textit{Results.}}—We consider in turn the case of symmetry restoration, for degenerate minima, and then vacuum decay between false and true vacua. Movies of the two processes can be found in \cite{movie}.
\vskip 5pt
\noindent
\textit{Symmetry restoration.}— We start by studying the case with $g=0$, where $\Delta V=0$ and, thus, the potential has two degenerate vacua at $\phi_{\pm}=\pm\eta$. In the left panel of Fig. \ref{fig:panel_1} we plot the potential, and in the right panel the evolution of the scalar field at a finite coordinate distance from the black hole, $r=2M$. While asymptotically far from the black hole the field oscillates around the negative minimum $\phi_{-}$, at smaller radii the field evolves over time. Initially it remains around $\phi_-$, but the amplitude of the oscillations grows due to the increased density resulting from the accretion. The scalar field starts to probe the quartic part of the potential, and when the cloud has become dense enough, the field transcends the barrier and reaches the other vacuum state at $\phi_+=+\eta$.

We can derive conditions on the timescale for the transition using the following argument. The field $\phi$ is at its maximum, when its time derivative is zero. The energy density is, then approximately given by
\begin{equation}
    \rho \approx V(\phi) + \frac{k^2}{2}(\phi+\eta)^2
\end{equation}
where $k\approx \mu$ is the typical inverse length scale of the gradient.
When the field reaches the positive vacuum, at $\phi_+=\eta$, the gradient term is the only contribution to the energy density, so that $\rho\approx 2k^2\eta^2 = 4\lambda \eta^4$.
We choose the asymptotic, background, value of the density to be $\rho_0\approx 0.03\lambda\eta^4$, which means the density near the black hole has to grow by $\approx 100$ times to reach $\rho\approx  4\lambda \eta^4$. The timescale for the transition is dependent on the accretion rate; extracting $\dot{\rho}\approx 10^{-3} M^{-3}$ from our simulations, we find that symmetry should be restored at $\Delta t\approx (\rho - \rho_0)/\dot{\rho} \approx 400 M$, which is consistent with what we observe in Fig. \ref{fig:panel_1}. After this state is reached the mean of the field near the black hole $\langle\phi\rangle_\mathrm{bh}$ appears to remain roughly constant, it now effectively evolves in a symmetric potential.

\begin{figure}[t!]
    \centering
    \href{https://youtu.be/c2vcNoX8fVs}{
    \includegraphics[width=\linewidth]{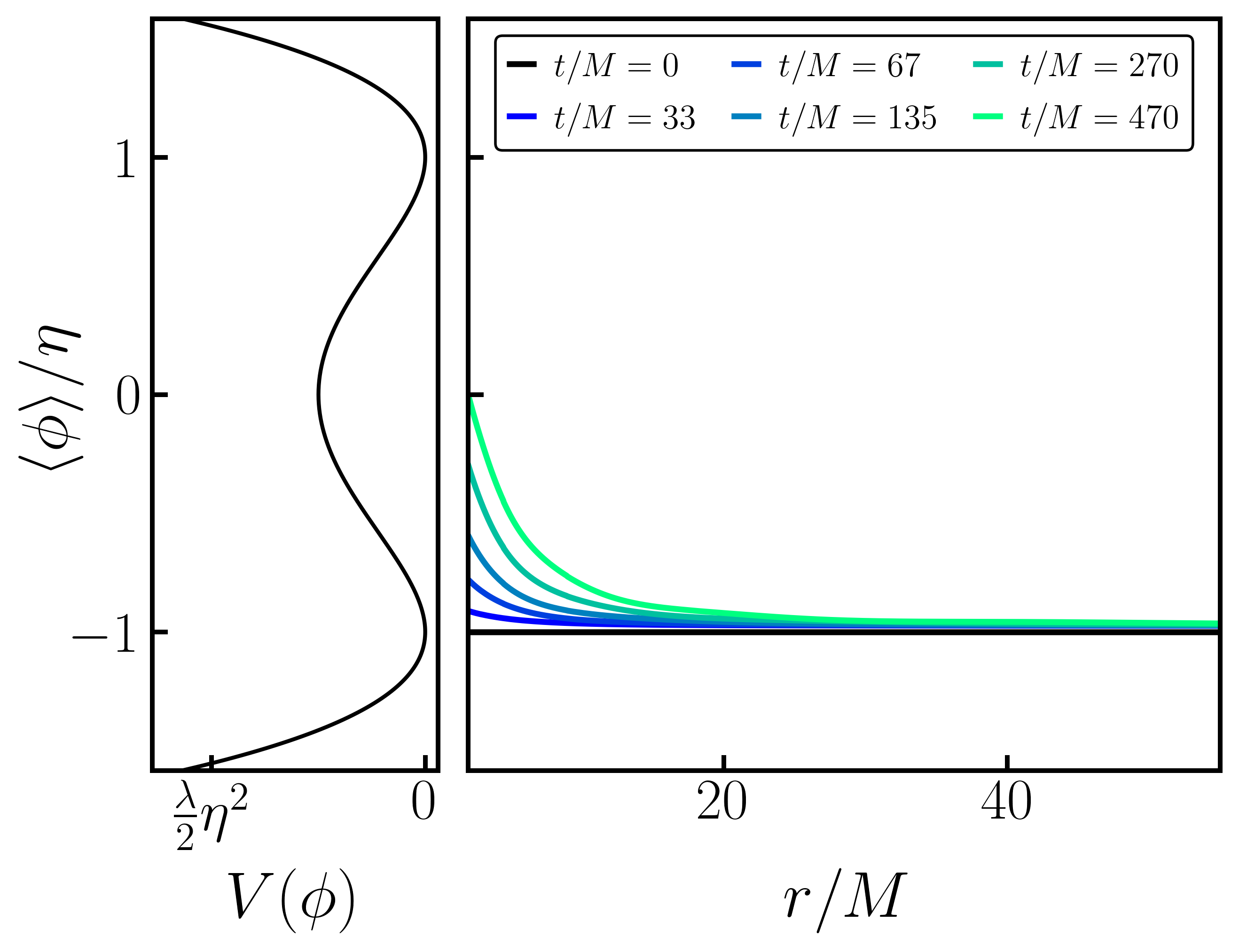}}
    \caption{Evolution of the mean scalar field radial profile over time. Symmetry is restored near the black hole horizon as $\phimeanbh=0$, and the field remains pinned down at $\phimeanin=-\eta$ asymptotically. The restoration phase slowly grows outwards as the black hole accretes more scalar field. }
    \label{fig:panel_r1}
\end{figure}

It is instructive to focus on the radial profiles of the mean value $\phimean$, which we plot in Fig. \ref{fig:panel_r1} at different times. Initially the scalar field oscillates around the negative minimum everywhere, so that $\phimean=-\eta$ for all radii. The scalar amplitude grows near the black hole due to the accretion, so that the field transcends the barrier near the horizon and starts to oscillate periodically over the full double-well potential. Hence, $\phimeanbh=0$ and symmetry is restored at that radius. However, the effects of gradients and asymptotic boundary conditions play an important role. At large radial distances from the black hole, the field remains oscillating around $\phimeanin=-\eta$, so that adjacent values of the field (in space) experience a restoring force towards the original minimum. Since the symmetry restored phase requires the field to reach a certain density, it will only extend out as far as this density is reached. Hence, the timescale for symmetry restoration will strongly depend on the accretion rate. From Fig. \ref{fig:panel_r1} we can infer that the symmetry restored phase propagates outwards, into the symmetry broken phase. This is a slow process, driven by the build up of the scalar matter via accretion, and so is different from the well-known expanding bubble of a new phase that is nucleated when there is a vacuum decay, which we will see in the following section.

What is the end state of this process? The density will only continue to increase whilst there is an asymptotic reservoir of the scalar feeding the growth. So only in the infinite accretion limit would the field fully restore symmetry. At some point the reservoir may be depleted and the cloud will then gradually decay into the black hole (a process that can be extremely long on astrophysical timescales for light scalars \cite{Barranco:2012qs}).

\vskip 5pt
\noindent
\textit{Vacuum decay.}— We now study how black holes can catalyze vacuum decay. To do so, we vary $g$ in Eqn. \eqref{eq:scalar_potential}, breaking the degeneracy between the minima by introducing a difference in vacuum potential energy densities $\Delta V$. If $g<0$, the energy difference between the minima is negative $\Delta V <0$, and $\phi_-$ and $\phi_+$ correspond to a false (metastable) and a true vacuum, respectively. This means that, as opposed to the degenerate case we have discussed before, there is now a preference for the field to decay to the lower energy state at $\phi_+$. 

\begin{figure}[t!]
    \centering
    \href{https://youtu.be/c2vcNoX8fVs}{
    \includegraphics[width=\linewidth]{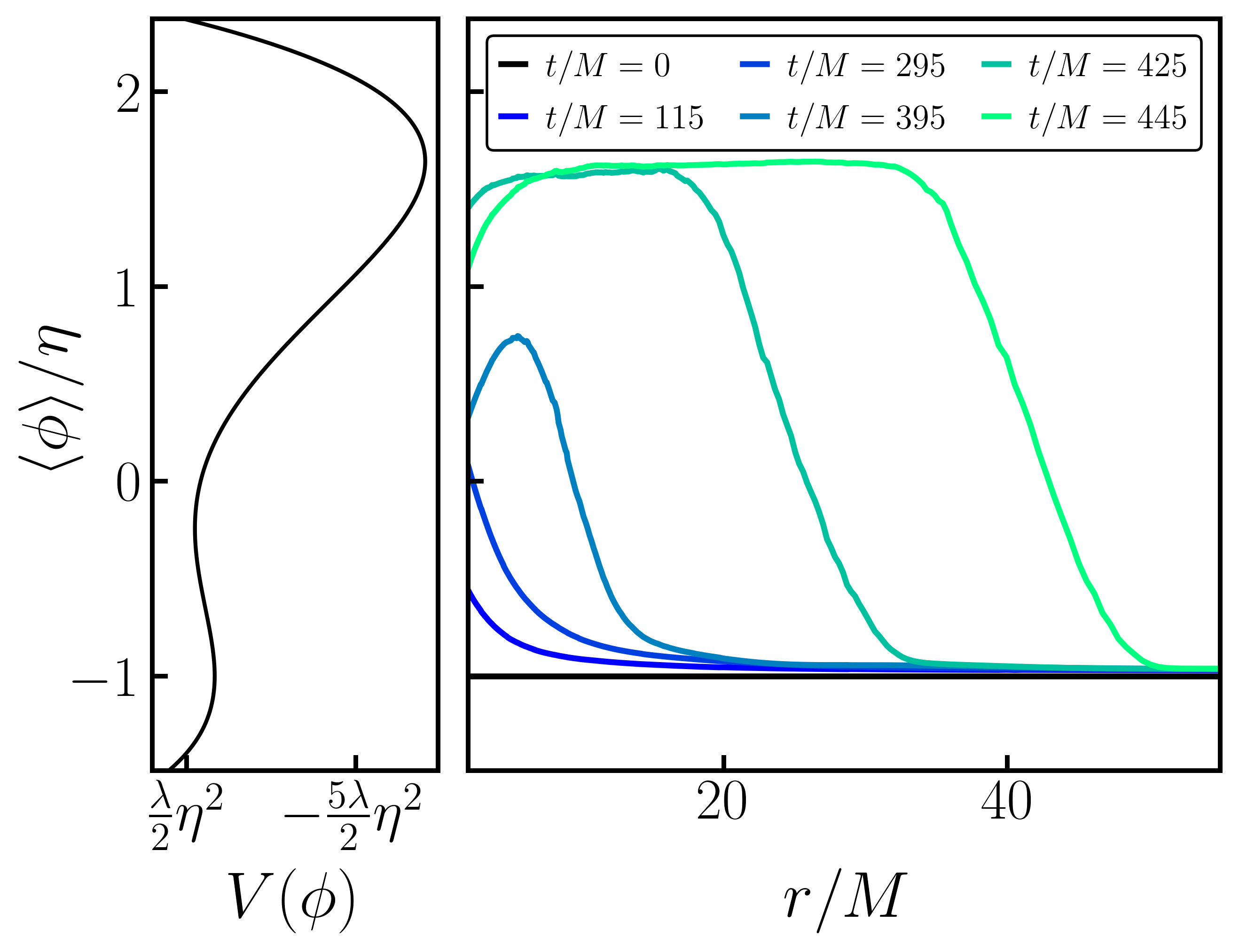}}
    \caption{Evolution of the mean $\phimean$ over time. The dynamics near the black hole triggers the nucleation of a true vacuum bubble that expands at ultrarelativistic speeds.}
    \label{fig:panel_r2}
\end{figure}

\begin{figure}[t!]
    \centering
    \href{https://youtu.be/c2vcNoX8fVs}{
    \includegraphics[width=\linewidth]{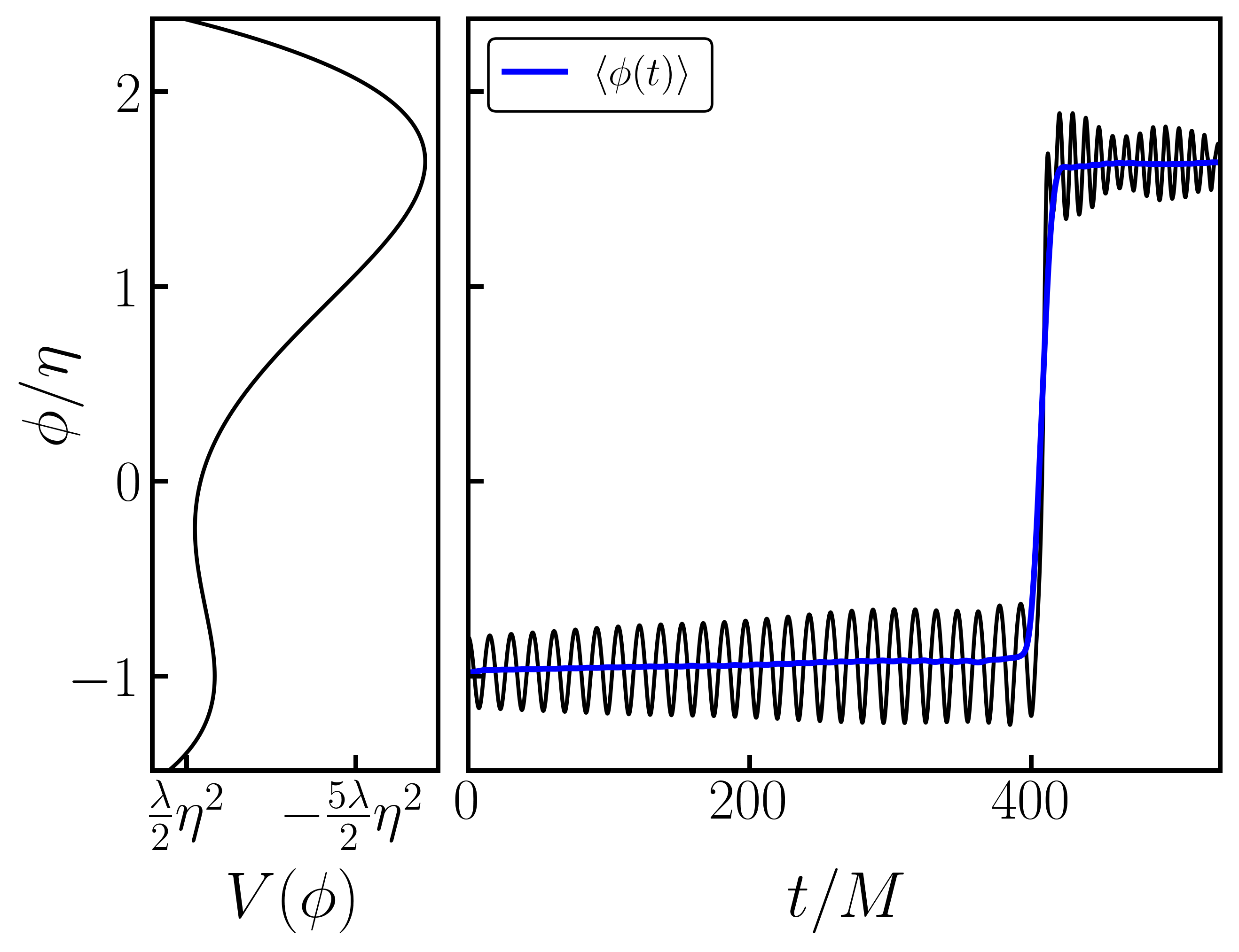}}
    \caption{Vacuum decay case, showing the value of the scalar field at $r=20M$ (black line). In blue we plot the mean value of the scalar field $\langle\phi(t)\rangle$, which transitions from $\phi_-$ to $\phi_+$.}
    \label{fig:panel_2}
\end{figure}

The process for $g=-0.04 M^{-2}$ is illustrated in Fig. \ref{fig:panel_r2}, which shows how the profile of the time averaged scalar field $\phimean$ changes over time. Close to the black hole, accretion drives the growth of the amplitude of the scalar field until it transcends the barrier and starts exploring the true vacuum $\phi_+$ by forming a true vacuum bubble. When this bubble is large enough so that the energy difference between vacua can overcome the wall tension, it is more favourable for the bubble to expand radially outwards. 
The energy difference $\Delta V$ is transferred to the bubble wall velocity, so that it quickly accelerates and expands at ultrarelativistic speeds. Gradients in the field once again play a crucial role. In this case, as opposed to keeping the field around the false minimum, they are responsible for dragging the rest of the field down to the true minimum, see Fig. \ref{fig:panel_2}. Whilst this is similar to how vacuum bubbles that formed via quantum tunneling propagate \cite{Coleman:1977py,Callan:1977pt,Coleman:1980aw},  the timescale for the decay here is completely deterministic. In our classical scenario, the associated timescale is given by the initial conditions around the black hole, the details of the potential and the accretion rate (which itself depends on the ratio $\mu M$). 

So what is the end state of this process?
As the true vacuum bubble expands, it drags down the field in the false vacuum, which remains oscillating around the true minimum $\phi_+$, as shown in Fig. \ref{fig:panel_2}. More interestingly, very close the black hole the symmetry restoration mechanism remains at play: the scalar field may sweep the whole potential, revisiting the false vacuum if the cloud density is large enough, see Fig. \ref{fig:panel_r2}. It would be interesting to study the possibility that this process could trigger the nucleation of a separate de Sitter universe when including backreaction, as observed in upward transitions of flyover vacuum decay \cite{Blanco-Pillado:2019xny}. If, on the other hand, the field is decaying to an AdS vacuum $V(\phi_+)<0$, the universe would transition to a collapsing phase that should eventually result in a Big Crunch, (although see also \cite{Cespedes:2020xpn} where scalar dynamics play a role). So although in this work we have ignored the effects of backreaction, this new vacuum decay mechanism triggered by black holes could have important gravitational consequences in the universe.

\vskip 10pt
\noindent
\textbf{\textit{Discussion.}}— In this work we have proposed a novel mechanism by which black holes can alter the vacuum structure of the universe. The presence of compact objects leads to the formation of overdensities in scalar fields in their vicinity via accretion, which translate to large-amplitude oscillations in the potential. In the case of a scalar field embedded in a symmetry broken phase, these oscillations can transcend the potential barrier separating different vacua. As a result, shells of symmetry restored phase may be created around black holes. In the case in which the field is in a false vacuum, this mechanism can also trigger vacuum decay near the black hole, forming a true vacuum bubble that propagates outwards at ultrarelativistic speeds and changes the vacuum structure even far away from the black hole.

Whilst here we focus on the case of $\mu M \sim 0.5$ (with $\mu^2 = 2\lambda \eta^2$), this process should be relevant for scalar fields and black holes whenever $\mu M \gtrsim 1$ -- the accretion clouds that we study are suppressed in the case of very light scalars \cite{Hui:2019aqm}. That is, for
\begin{equation}\label{eq:pred}
    \left(\frac{\mu}{10^{-11}\mathrm{eV}}\right)\left(\frac{M}{M_\odot}\right) \gtrsim 1\,.
\end{equation}
The decay rate of this classical mechanism is completely deterministic and depends on the accretion rate of the black hole. Using the growth rate for the case that we have simulated, the approximate timescale for symmetry restoration was
\begin{equation}
\Delta t\approx 10^{-3}\,  \mathrm{sec}\, \left(\frac{M}{M_\odot}\right)
\end{equation}
The growth rate should be slower for smaller $\mu$, lower initial densities and higher values of $\eta$, as well as other overdensity scenarios, such as binary black holes, which can enhance the local density \cite{Bamber:2022pbs,Aurrekoetxea:2023jwk} and thus may enhance the transition rate.
Since we neglect backreaction, our simulations permit us to rescale quantities and describe any case for which $\mu M \sim 0.5$ and
\begin{equation}\label{eq:scales}
    \left(\frac{\rho_0}{10 \, {\rm GeV / cm^{3}}}\right)^{-1/2} \left(\frac{\eta}{\mathrm{GeV}}\right)^{-1} \left(\frac{M}{M_\odot}\right)  \sim 1\,,
\end{equation}
(assuming that the density remains small relative to the curvature scale of the BH).
We conclude that the presence of black holes could have implications for a range of symmetry breaking scales $\eta$ and masses $\mu$, including in the early universe. 

\clearpage

\vskip 10pt
\noindent
\textbf{\textit{Acknowledgements.}}—  We thank the GRChombo collaboration (\href{www.grchombo.org}{www.grchombo.org}) for their support and code development work. JM acknowledges funding from STFC. JCA acknowledges funding from the Beecroft Trust and The Queen’s College via an extraordinary Junior Research Fellowship (eJRF). KC acknowledges funding from the UKRI Ernest Rutherford Fellowship (grant number ST/V003240/1) and STFC Research Grant ST/X000931/1 (Astronomy at Queen Mary 2023-2026). PGF acknowledges support from STFC and the Beecroft Trust.

Part of this work was performed using the DiRAC@Durham facility managed by the Institute for Computational Cosmology on behalf of the STFC DiRAC HPC Facility (www.dirac.ac.uk) under  DiRAC RAC13 Grant ACTP238 and DiRAC RAC15 Grant ACTP316. The equipment was funded by BEIS capital funding via STFC capital grants ST/P002293/1, ST/R002371/1 and ST/S002502/1, Durham University and STFC operations
grant ST/R000832/1.
This work also used the DiRAC Data Intensive service at Leicester, operated by the University of Leicester IT Services, which forms part of the STFC DiRAC HPC Facility (www.dirac.ac.uk). The equipment was funded by BEIS capital funding via STFC capital grants ST/K000373/1 and ST/R002363/1 and STFC DiRAC Operations Grant ST/R001014/1. DiRAC is part of the National e-Infrastructure.

\bibliography{refs.bib}

\begin{thebibliography}{60}%
\makeatletter
\providecommand \@ifxundefined [1]{%
 \@ifx{#1\undefined}
}%
\providecommand \@ifnum [1]{%
 \ifnum #1\expandafter \@firstoftwo
 \else \expandafter \@secondoftwo
 \fi
}%
\providecommand \@ifx [1]{%
 \ifx #1\expandafter \@firstoftwo
 \else \expandafter \@secondoftwo
 \fi
}%
\providecommand \natexlab [1]{#1}%
\providecommand \enquote  [1]{``#1''}%
\providecommand \bibnamefont  [1]{#1}%
\providecommand \bibfnamefont [1]{#1}%
\providecommand \citenamefont [1]{#1}%
\providecommand \href@noop [0]{\@secondoftwo}%
\providecommand \href [0]{\begingroup \@sanitize@url \@href}%
\providecommand \@href[1]{\@@startlink{#1}\@@href}%
\providecommand \@@href[1]{\endgroup#1\@@endlink}%
\providecommand \@sanitize@url [0]{\catcode `\\12\catcode `\$12\catcode `\&12\catcode `\#12\catcode `\^12\catcode `\_12\catcode `\%12\relax}%
\providecommand \@@startlink[1]{}%
\providecommand \@@endlink[0]{}%
\providecommand \url  [0]{\begingroup\@sanitize@url \@url }%
\providecommand \@url [1]{\endgroup\@href {#1}{\urlprefix }}%
\providecommand \urlprefix  [0]{URL }%
\providecommand \Eprint [0]{\href }%
\providecommand \doibase [0]{http://dx.doi.org/}%
\providecommand \selectlanguage [0]{\@gobble}%
\providecommand \bibinfo  [0]{\@secondoftwo}%
\providecommand \bibfield  [0]{\@secondoftwo}%
\providecommand \translation [1]{[#1]}%
\providecommand \BibitemOpen [0]{}%
\providecommand \bibitemStop [0]{}%
\providecommand \bibitemNoStop [0]{.\EOS\space}%
\providecommand \EOS [0]{\spacefactor3000\relax}%
\providecommand \BibitemShut  [1]{\csname bibitem#1\endcsname}%
\let\auto@bib@innerbib\@empty
\bibitem [{\citenamefont {Peter}\ and\ \citenamefont {Uzan}(2013)}]{Peter:2013avv}%
  \BibitemOpen
  \bibfield  {author} {\bibinfo {author} {\bibfnamefont {P.}~\bibnamefont {Peter}}\ and\ \bibinfo {author} {\bibfnamefont {J.-P.}\ \bibnamefont {Uzan}},\ }\href@noop {} {\emph {\bibinfo {title} {{Primordial Cosmology}}}},\ Oxford Graduate Texts\ (\bibinfo  {publisher} {Oxford University Press},\ \bibinfo {year} {2013})\BibitemShut {NoStop}%
\bibitem [{\citenamefont {Coleman}\ and\ \citenamefont {Weinberg}(1973)}]{Coleman:1973jx}%
  \BibitemOpen
  \bibfield  {author} {\bibinfo {author} {\bibfnamefont {S.~R.}\ \bibnamefont {Coleman}}\ and\ \bibinfo {author} {\bibfnamefont {E.~J.}\ \bibnamefont {Weinberg}},\ }\href {\doibase 10.1103/PhysRevD.7.1888} {\bibfield  {journal} {\bibinfo  {journal} {Phys. Rev. D}\ }\textbf {\bibinfo {volume} {7}},\ \bibinfo {pages} {1888} (\bibinfo {year} {1973})}\BibitemShut {NoStop}%
\bibitem [{\citenamefont {Coleman}(1977)}]{Coleman:1977py}%
  \BibitemOpen
  \bibfield  {author} {\bibinfo {author} {\bibfnamefont {S.~R.}\ \bibnamefont {Coleman}},\ }\href {\doibase 10.1103/PhysRevD.16.1248} {\bibfield  {journal} {\bibinfo  {journal} {Phys. Rev. D}\ }\textbf {\bibinfo {volume} {15}},\ \bibinfo {pages} {2929} (\bibinfo {year} {1977})},\ \bibinfo {note} {[Erratum: Phys.Rev.D 16, 1248 (1977)]}\BibitemShut {NoStop}%
\bibitem [{\citenamefont {Callan}\ and\ \citenamefont {Coleman}(1977)}]{Callan:1977pt}%
  \BibitemOpen
  \bibfield  {author} {\bibinfo {author} {\bibfnamefont {C.~G.}\ \bibnamefont {Callan}, \bibfnamefont {Jr.}}\ and\ \bibinfo {author} {\bibfnamefont {S.~R.}\ \bibnamefont {Coleman}},\ }\href {\doibase 10.1103/PhysRevD.16.1762} {\bibfield  {journal} {\bibinfo  {journal} {Phys. Rev. D}\ }\textbf {\bibinfo {volume} {16}},\ \bibinfo {pages} {1762} (\bibinfo {year} {1977})}\BibitemShut {NoStop}%
\bibitem [{\citenamefont {Coleman}\ and\ \citenamefont {De~Luccia}(1980)}]{Coleman:1980aw}%
  \BibitemOpen
  \bibfield  {author} {\bibinfo {author} {\bibfnamefont {S.~R.}\ \bibnamefont {Coleman}}\ and\ \bibinfo {author} {\bibfnamefont {F.}~\bibnamefont {De~Luccia}},\ }\href {\doibase 10.1103/PhysRevD.21.3305} {\bibfield  {journal} {\bibinfo  {journal} {Phys. Rev. D}\ }\textbf {\bibinfo {volume} {21}},\ \bibinfo {pages} {3305} (\bibinfo {year} {1980})}\BibitemShut {NoStop}%
\bibitem [{\citenamefont {Hawking}\ and\ \citenamefont {Moss}(1982)}]{Hawking:1981fz}%
  \BibitemOpen
  \bibfield  {author} {\bibinfo {author} {\bibfnamefont {S.~W.}\ \bibnamefont {Hawking}}\ and\ \bibinfo {author} {\bibfnamefont {I.~G.}\ \bibnamefont {Moss}},\ }\href {\doibase 10.1016/0370-2693(82)90946-7} {\bibfield  {journal} {\bibinfo  {journal} {Phys. Lett. B}\ }\textbf {\bibinfo {volume} {110}},\ \bibinfo {pages} {35} (\bibinfo {year} {1982})}\BibitemShut {NoStop}%
\bibitem [{\citenamefont {Berezin}\ \emph {et~al.}(1988)\citenamefont {Berezin}, \citenamefont {Kuzmin},\ and\ \citenamefont {Tkachev}}]{Berezin:1987ea}%
  \BibitemOpen
  \bibfield  {author} {\bibinfo {author} {\bibfnamefont {V.~A.}\ \bibnamefont {Berezin}}, \bibinfo {author} {\bibfnamefont {V.~A.}\ \bibnamefont {Kuzmin}}, \ and\ \bibinfo {author} {\bibfnamefont {I.~I.}\ \bibnamefont {Tkachev}},\ }\href {\doibase 10.1016/0370-2693(88)90672-7} {\bibfield  {journal} {\bibinfo  {journal} {Phys. Lett. B}\ }\textbf {\bibinfo {volume} {207}},\ \bibinfo {pages} {397} (\bibinfo {year} {1988})}\BibitemShut {NoStop}%
\bibitem [{\citenamefont {Hiscock}(1987)}]{Hiscock:1987hn}%
  \BibitemOpen
  \bibfield  {author} {\bibinfo {author} {\bibfnamefont {W.~A.}\ \bibnamefont {Hiscock}},\ }\href {\doibase 10.1103/PhysRevD.35.1161} {\bibfield  {journal} {\bibinfo  {journal} {Phys. Rev. D}\ }\textbf {\bibinfo {volume} {35}},\ \bibinfo {pages} {1161} (\bibinfo {year} {1987})}\BibitemShut {NoStop}%
\bibitem [{\citenamefont {Arnold}(1990)}]{Arnold:1989cq}%
  \BibitemOpen
  \bibfield  {author} {\bibinfo {author} {\bibfnamefont {P.~B.}\ \bibnamefont {Arnold}},\ }\href {\doibase 10.1016/0550-3213(90)90243-7} {\bibfield  {journal} {\bibinfo  {journal} {Nucl. Phys. B}\ }\textbf {\bibinfo {volume} {346}},\ \bibinfo {pages} {160} (\bibinfo {year} {1990})}\BibitemShut {NoStop}%
\bibitem [{\citenamefont {Berezin}\ \emph {et~al.}(1991)\citenamefont {Berezin}, \citenamefont {Kuzmin},\ and\ \citenamefont {Tkachev}}]{Berezin:1990qs}%
  \BibitemOpen
  \bibfield  {author} {\bibinfo {author} {\bibfnamefont {V.~A.}\ \bibnamefont {Berezin}}, \bibinfo {author} {\bibfnamefont {V.~A.}\ \bibnamefont {Kuzmin}}, \ and\ \bibinfo {author} {\bibfnamefont {I.~I.}\ \bibnamefont {Tkachev}},\ }\href {\doibase 10.1103/PhysRevD.43.R3112} {\bibfield  {journal} {\bibinfo  {journal} {Phys. Rev. D}\ }\textbf {\bibinfo {volume} {43}},\ \bibinfo {pages} {3112} (\bibinfo {year} {1991})}\BibitemShut {NoStop}%
\bibitem [{\citenamefont {Gregory}\ \emph {et~al.}(2014)\citenamefont {Gregory}, \citenamefont {Moss},\ and\ \citenamefont {Withers}}]{Gregory:2013hja}%
  \BibitemOpen
  \bibfield  {author} {\bibinfo {author} {\bibfnamefont {R.}~\bibnamefont {Gregory}}, \bibinfo {author} {\bibfnamefont {I.~G.}\ \bibnamefont {Moss}}, \ and\ \bibinfo {author} {\bibfnamefont {B.}~\bibnamefont {Withers}},\ }\href {\doibase 10.1007/JHEP03(2014)081} {\bibfield  {journal} {\bibinfo  {journal} {JHEP}\ }\textbf {\bibinfo {volume} {03}},\ \bibinfo {pages} {081} (\bibinfo {year} {2014})},\ \Eprint {http://arxiv.org/abs/1401.0017} {arXiv:1401.0017 [hep-th]} \BibitemShut {NoStop}%
\bibitem [{\citenamefont {Burda}\ \emph {et~al.}(2015{\natexlab{a}})\citenamefont {Burda}, \citenamefont {Gregory},\ and\ \citenamefont {Moss}}]{Burda:2015yfa}%
  \BibitemOpen
  \bibfield  {author} {\bibinfo {author} {\bibfnamefont {P.}~\bibnamefont {Burda}}, \bibinfo {author} {\bibfnamefont {R.}~\bibnamefont {Gregory}}, \ and\ \bibinfo {author} {\bibfnamefont {I.}~\bibnamefont {Moss}},\ }\href {\doibase 10.1007/JHEP08(2015)114} {\bibfield  {journal} {\bibinfo  {journal} {JHEP}\ }\textbf {\bibinfo {volume} {08}},\ \bibinfo {pages} {114} (\bibinfo {year} {2015}{\natexlab{a}})},\ \Eprint {http://arxiv.org/abs/1503.07331} {arXiv:1503.07331 [hep-th]} \BibitemShut {NoStop}%
\bibitem [{\citenamefont {Burda}\ \emph {et~al.}(2015{\natexlab{b}})\citenamefont {Burda}, \citenamefont {Gregory},\ and\ \citenamefont {Moss}}]{Burda:2015isa}%
  \BibitemOpen
  \bibfield  {author} {\bibinfo {author} {\bibfnamefont {P.}~\bibnamefont {Burda}}, \bibinfo {author} {\bibfnamefont {R.}~\bibnamefont {Gregory}}, \ and\ \bibinfo {author} {\bibfnamefont {I.}~\bibnamefont {Moss}},\ }\href {\doibase 10.1103/PhysRevLett.115.071303} {\bibfield  {journal} {\bibinfo  {journal} {Phys. Rev. Lett.}\ }\textbf {\bibinfo {volume} {115}},\ \bibinfo {pages} {071303} (\bibinfo {year} {2015}{\natexlab{b}})},\ \Eprint {http://arxiv.org/abs/1501.04937} {arXiv:1501.04937 [hep-th]} \BibitemShut {NoStop}%
\bibitem [{\citenamefont {Burda}\ \emph {et~al.}(2016)\citenamefont {Burda}, \citenamefont {Gregory},\ and\ \citenamefont {Moss}}]{Burda:2016mou}%
  \BibitemOpen
  \bibfield  {author} {\bibinfo {author} {\bibfnamefont {P.}~\bibnamefont {Burda}}, \bibinfo {author} {\bibfnamefont {R.}~\bibnamefont {Gregory}}, \ and\ \bibinfo {author} {\bibfnamefont {I.}~\bibnamefont {Moss}},\ }\href {\doibase 10.1007/JHEP06(2016)025} {\bibfield  {journal} {\bibinfo  {journal} {JHEP}\ }\textbf {\bibinfo {volume} {06}},\ \bibinfo {pages} {025} (\bibinfo {year} {2016})},\ \Eprint {http://arxiv.org/abs/1601.02152} {arXiv:1601.02152 [hep-th]} \BibitemShut {NoStop}%
\bibitem [{\citenamefont {Mukaida}\ and\ \citenamefont {Yamada}(2017)}]{Mukaida:2017bgd}%
  \BibitemOpen
  \bibfield  {author} {\bibinfo {author} {\bibfnamefont {K.}~\bibnamefont {Mukaida}}\ and\ \bibinfo {author} {\bibfnamefont {M.}~\bibnamefont {Yamada}},\ }\href {\doibase 10.1103/PhysRevD.96.103514} {\bibfield  {journal} {\bibinfo  {journal} {Phys. Rev. D}\ }\textbf {\bibinfo {volume} {96}},\ \bibinfo {pages} {103514} (\bibinfo {year} {2017})},\ \Eprint {http://arxiv.org/abs/1706.04523} {arXiv:1706.04523 [hep-th]} \BibitemShut {NoStop}%
\bibitem [{\citenamefont {Oshita}\ \emph {et~al.}(2019)\citenamefont {Oshita}, \citenamefont {Yamada},\ and\ \citenamefont {Yamaguchi}}]{Oshita:2018ptr}%
  \BibitemOpen
  \bibfield  {author} {\bibinfo {author} {\bibfnamefont {N.}~\bibnamefont {Oshita}}, \bibinfo {author} {\bibfnamefont {M.}~\bibnamefont {Yamada}}, \ and\ \bibinfo {author} {\bibfnamefont {M.}~\bibnamefont {Yamaguchi}},\ }\href {\doibase 10.1016/j.physletb.2019.02.032} {\bibfield  {journal} {\bibinfo  {journal} {Phys. Lett. B}\ }\textbf {\bibinfo {volume} {791}},\ \bibinfo {pages} {149} (\bibinfo {year} {2019})},\ \Eprint {http://arxiv.org/abs/1808.01382} {arXiv:1808.01382 [gr-qc]} \BibitemShut {NoStop}%
\bibitem [{\citenamefont {Oshita}\ \emph {et~al.}(2020)\citenamefont {Oshita}, \citenamefont {Ueda},\ and\ \citenamefont {Yamaguchi}}]{Oshita:2019jan}%
  \BibitemOpen
  \bibfield  {author} {\bibinfo {author} {\bibfnamefont {N.}~\bibnamefont {Oshita}}, \bibinfo {author} {\bibfnamefont {K.}~\bibnamefont {Ueda}}, \ and\ \bibinfo {author} {\bibfnamefont {M.}~\bibnamefont {Yamaguchi}},\ }\href {\doibase 10.1007/JHEP01(2020)015} {\bibfield  {journal} {\bibinfo  {journal} {JHEP}\ }\textbf {\bibinfo {volume} {01}},\ \bibinfo {pages} {015} (\bibinfo {year} {2020})},\ \bibinfo {note} {[Erratum: JHEP 10, 122 (2020)]},\ \Eprint {http://arxiv.org/abs/1909.01378} {arXiv:1909.01378 [hep-th]} \BibitemShut {NoStop}%
\bibitem [{\citenamefont {Hui}(2021)}]{Hui:2021tkt}%
  \BibitemOpen
  \bibfield  {author} {\bibinfo {author} {\bibfnamefont {L.}~\bibnamefont {Hui}},\ }\href {\doibase 10.1146/annurev-astro-120920-010024} {\bibfield  {journal} {\bibinfo  {journal} {Ann. Rev. Astron. Astrophys.}\ }\textbf {\bibinfo {volume} {59}},\ \bibinfo {pages} {247} (\bibinfo {year} {2021})},\ \Eprint {http://arxiv.org/abs/2101.11735} {arXiv:2101.11735 [astro-ph.CO]} \BibitemShut {NoStop}%
\bibitem [{\citenamefont {Marsden}\ \emph {et~al.}(2024)\citenamefont {Marsden}, \citenamefont {Aurrekoetxea}, \citenamefont {Clough},\ and\ \citenamefont {Ferreira}}]{Marsden:SIpaper}%
  \BibitemOpen
  \bibfield  {author} {\bibinfo {author} {\bibfnamefont {J.}~\bibnamefont {Marsden}}, \bibinfo {author} {\bibfnamefont {J.~C.}\ \bibnamefont {Aurrekoetxea}}, \bibinfo {author} {\bibfnamefont {K.}~\bibnamefont {Clough}}, \ and\ \bibinfo {author} {\bibfnamefont {P.~G.}\ \bibnamefont {Ferreira}},\ }\href@noop {} {\  (\bibinfo {year} {2024})},\ \Eprint {http://arxiv.org/abs/in prep} {arXiv:in prep [gr-qc]} \BibitemShut {NoStop}%
\bibitem [{\citenamefont {Shapiro}\ and\ \citenamefont {Paschalidis}(2014)}]{Shapiro:2014oha}%
  \BibitemOpen
  \bibfield  {author} {\bibinfo {author} {\bibfnamefont {S.~L.}\ \bibnamefont {Shapiro}}\ and\ \bibinfo {author} {\bibfnamefont {V.}~\bibnamefont {Paschalidis}},\ }\href {\doibase 10.1103/PhysRevD.89.023506} {\bibfield  {journal} {\bibinfo  {journal} {Phys. Rev. D}\ }\textbf {\bibinfo {volume} {89}},\ \bibinfo {pages} {023506} (\bibinfo {year} {2014})},\ \Eprint {http://arxiv.org/abs/1402.0005} {arXiv:1402.0005 [astro-ph.CO]} \BibitemShut {NoStop}%
\bibitem [{\citenamefont {Berezhiani}\ \emph {et~al.}(2023)\citenamefont {Berezhiani}, \citenamefont {Cintia}, \citenamefont {De~Luca},\ and\ \citenamefont {Khoury}}]{Berezhiani:2023vlo}%
  \BibitemOpen
  \bibfield  {author} {\bibinfo {author} {\bibfnamefont {L.}~\bibnamefont {Berezhiani}}, \bibinfo {author} {\bibfnamefont {G.}~\bibnamefont {Cintia}}, \bibinfo {author} {\bibfnamefont {V.}~\bibnamefont {De~Luca}}, \ and\ \bibinfo {author} {\bibfnamefont {J.}~\bibnamefont {Khoury}},\ }\href@noop {} {\  (\bibinfo {year} {2023})},\ \Eprint {http://arxiv.org/abs/2311.07672} {arXiv:2311.07672 [astro-ph.CO]} \BibitemShut {NoStop}%
\bibitem [{\citenamefont {Kadota}\ \emph {et~al.}(2024)\citenamefont {Kadota}, \citenamefont {Kim}, \citenamefont {Ko},\ and\ \citenamefont {Yang}}]{Kadota:2023wlm}%
  \BibitemOpen
  \bibfield  {author} {\bibinfo {author} {\bibfnamefont {K.}~\bibnamefont {Kadota}}, \bibinfo {author} {\bibfnamefont {J.~H.}\ \bibnamefont {Kim}}, \bibinfo {author} {\bibfnamefont {P.}~\bibnamefont {Ko}}, \ and\ \bibinfo {author} {\bibfnamefont {X.-Y.}\ \bibnamefont {Yang}},\ }\href {\doibase 10.1103/PhysRevD.109.015022} {\bibfield  {journal} {\bibinfo  {journal} {Phys. Rev. D}\ }\textbf {\bibinfo {volume} {109}},\ \bibinfo {pages} {015022} (\bibinfo {year} {2024})},\ \Eprint {http://arxiv.org/abs/2306.10828} {arXiv:2306.10828 [hep-ph]} \BibitemShut {NoStop}%
\bibitem [{\citenamefont {Alonso-\'Alvarez}\ \emph {et~al.}(2024)\citenamefont {Alonso-\'Alvarez}, \citenamefont {Cline},\ and\ \citenamefont {Dewar}}]{Alonso-Alvarez:2024gdz}%
  \BibitemOpen
  \bibfield  {author} {\bibinfo {author} {\bibfnamefont {G.}~\bibnamefont {Alonso-\'Alvarez}}, \bibinfo {author} {\bibfnamefont {J.~M.}\ \bibnamefont {Cline}}, \ and\ \bibinfo {author} {\bibfnamefont {C.}~\bibnamefont {Dewar}},\ }\href@noop {} {\  (\bibinfo {year} {2024})},\ \Eprint {http://arxiv.org/abs/2401.14450} {arXiv:2401.14450 [astro-ph.CO]} \BibitemShut {NoStop}%
\bibitem [{\citenamefont {Boudon}\ \emph {et~al.}(2024)\citenamefont {Boudon}, \citenamefont {Brax}, \citenamefont {Valageas},\ and\ \citenamefont {Wong}}]{Boudon:2023vzl}%
  \BibitemOpen
  \bibfield  {author} {\bibinfo {author} {\bibfnamefont {A.}~\bibnamefont {Boudon}}, \bibinfo {author} {\bibfnamefont {P.}~\bibnamefont {Brax}}, \bibinfo {author} {\bibfnamefont {P.}~\bibnamefont {Valageas}}, \ and\ \bibinfo {author} {\bibfnamefont {L.~K.}\ \bibnamefont {Wong}},\ }\href {\doibase 10.1103/PhysRevD.109.043504} {\bibfield  {journal} {\bibinfo  {journal} {Phys. Rev. D}\ }\textbf {\bibinfo {volume} {109}},\ \bibinfo {pages} {043504} (\bibinfo {year} {2024})},\ \Eprint {http://arxiv.org/abs/2305.18540} {arXiv:2305.18540 [astro-ph.CO]} \BibitemShut {NoStop}%
\bibitem [{\citenamefont {Feng}\ \emph {et~al.}(2022)\citenamefont {Feng}, \citenamefont {Parisi}, \citenamefont {Chen},\ and\ \citenamefont {Lin}}]{Feng:2021qkj}%
  \BibitemOpen
  \bibfield  {author} {\bibinfo {author} {\bibfnamefont {W.-X.}\ \bibnamefont {Feng}}, \bibinfo {author} {\bibfnamefont {A.}~\bibnamefont {Parisi}}, \bibinfo {author} {\bibfnamefont {C.-S.}\ \bibnamefont {Chen}}, \ and\ \bibinfo {author} {\bibfnamefont {F.-L.}\ \bibnamefont {Lin}},\ }\href {\doibase 10.1088/1475-7516/2022/08/032} {\bibfield  {journal} {\bibinfo  {journal} {JCAP}\ }\textbf {\bibinfo {volume} {08}},\ \bibinfo {pages} {032} (\bibinfo {year} {2022})},\ \Eprint {http://arxiv.org/abs/2112.05160} {arXiv:2112.05160 [astro-ph.HE]} \BibitemShut {NoStop}%
\bibitem [{\citenamefont {Alvarez}\ and\ \citenamefont {Yu}(2021)}]{Alvarez:2020fyo}%
  \BibitemOpen
  \bibfield  {author} {\bibinfo {author} {\bibfnamefont {G.}~\bibnamefont {Alvarez}}\ and\ \bibinfo {author} {\bibfnamefont {H.-B.}\ \bibnamefont {Yu}},\ }\href {\doibase 10.1103/PhysRevD.104.043013} {\bibfield  {journal} {\bibinfo  {journal} {Phys. Rev. D}\ }\textbf {\bibinfo {volume} {104}},\ \bibinfo {pages} {043013} (\bibinfo {year} {2021})},\ \Eprint {http://arxiv.org/abs/2012.15050} {arXiv:2012.15050 [hep-ph]} \BibitemShut {NoStop}%
\bibitem [{\citenamefont {Yoshino}\ and\ \citenamefont {Kodama}(2012)}]{Yoshino:2012kn}%
  \BibitemOpen
  \bibfield  {author} {\bibinfo {author} {\bibfnamefont {H.}~\bibnamefont {Yoshino}}\ and\ \bibinfo {author} {\bibfnamefont {H.}~\bibnamefont {Kodama}},\ }\href {\doibase 10.1143/PTP.128.153} {\bibfield  {journal} {\bibinfo  {journal} {Prog. Theor. Phys.}\ }\textbf {\bibinfo {volume} {128}},\ \bibinfo {pages} {153} (\bibinfo {year} {2012})},\ \Eprint {http://arxiv.org/abs/1203.5070} {arXiv:1203.5070 [gr-qc]} \BibitemShut {NoStop}%
\bibitem [{\citenamefont {Yoshino}\ and\ \citenamefont {Kodama}(2015)}]{Yoshino:2015nsa}%
  \BibitemOpen
  \bibfield  {author} {\bibinfo {author} {\bibfnamefont {H.}~\bibnamefont {Yoshino}}\ and\ \bibinfo {author} {\bibfnamefont {H.}~\bibnamefont {Kodama}},\ }\href {\doibase 10.1088/0264-9381/32/21/214001} {\bibfield  {journal} {\bibinfo  {journal} {Class. Quant. Grav.}\ }\textbf {\bibinfo {volume} {32}},\ \bibinfo {pages} {214001} (\bibinfo {year} {2015})},\ \Eprint {http://arxiv.org/abs/1505.00714} {arXiv:1505.00714 [gr-qc]} \BibitemShut {NoStop}%
\bibitem [{\citenamefont {Omiya}\ \emph {et~al.}(2023)\citenamefont {Omiya}, \citenamefont {Takahashi}, \citenamefont {Tanaka},\ and\ \citenamefont {Yoshino}}]{Omiya:2022gwu}%
  \BibitemOpen
  \bibfield  {author} {\bibinfo {author} {\bibfnamefont {H.}~\bibnamefont {Omiya}}, \bibinfo {author} {\bibfnamefont {T.}~\bibnamefont {Takahashi}}, \bibinfo {author} {\bibfnamefont {T.}~\bibnamefont {Tanaka}}, \ and\ \bibinfo {author} {\bibfnamefont {H.}~\bibnamefont {Yoshino}},\ }\href {\doibase 10.1088/1475-7516/2023/06/016} {\bibfield  {journal} {\bibinfo  {journal} {JCAP}\ }\textbf {\bibinfo {volume} {06}},\ \bibinfo {pages} {016} (\bibinfo {year} {2023})},\ \Eprint {http://arxiv.org/abs/2211.01949} {arXiv:2211.01949 [gr-qc]} \BibitemShut {NoStop}%
\bibitem [{\citenamefont {Omiya}\ \emph {et~al.}(2022)\citenamefont {Omiya}, \citenamefont {Takahashi},\ and\ \citenamefont {Tanaka}}]{Omiya:2022mwv}%
  \BibitemOpen
  \bibfield  {author} {\bibinfo {author} {\bibfnamefont {H.}~\bibnamefont {Omiya}}, \bibinfo {author} {\bibfnamefont {T.}~\bibnamefont {Takahashi}}, \ and\ \bibinfo {author} {\bibfnamefont {T.}~\bibnamefont {Tanaka}},\ }\href {\doibase 10.1093/ptep/ptac058} {\bibfield  {journal} {\bibinfo  {journal} {PTEP}\ }\textbf {\bibinfo {volume} {2022}},\ \bibinfo {pages} {043E03} (\bibinfo {year} {2022})},\ \Eprint {http://arxiv.org/abs/2201.04382} {arXiv:2201.04382 [gr-qc]} \BibitemShut {NoStop}%
\bibitem [{\citenamefont {Omiya}\ \emph {et~al.}(2021)\citenamefont {Omiya}, \citenamefont {Takahashi},\ and\ \citenamefont {Tanaka}}]{Omiya:2020vji}%
  \BibitemOpen
  \bibfield  {author} {\bibinfo {author} {\bibfnamefont {H.}~\bibnamefont {Omiya}}, \bibinfo {author} {\bibfnamefont {T.}~\bibnamefont {Takahashi}}, \ and\ \bibinfo {author} {\bibfnamefont {T.}~\bibnamefont {Tanaka}},\ }\href {\doibase 10.1093/ptep/ptab032} {\bibfield  {journal} {\bibinfo  {journal} {PTEP}\ }\textbf {\bibinfo {volume} {2021}},\ \bibinfo {pages} {043E02} (\bibinfo {year} {2021})},\ \Eprint {http://arxiv.org/abs/2012.03473} {arXiv:2012.03473 [gr-qc]} \BibitemShut {NoStop}%
\bibitem [{\citenamefont {East}\ and\ \citenamefont {Huang}(2022)}]{East:2022rsi}%
  \BibitemOpen
  \bibfield  {author} {\bibinfo {author} {\bibfnamefont {W.~E.}\ \bibnamefont {East}}\ and\ \bibinfo {author} {\bibfnamefont {J.}~\bibnamefont {Huang}},\ }\href {\doibase 10.1007/JHEP12(2022)089} {\bibfield  {journal} {\bibinfo  {journal} {JHEP}\ }\textbf {\bibinfo {volume} {12}},\ \bibinfo {pages} {089} (\bibinfo {year} {2022})},\ \Eprint {http://arxiv.org/abs/2206.12432} {arXiv:2206.12432 [hep-ph]} \BibitemShut {NoStop}%
\bibitem [{\citenamefont {East}(2022)}]{East:2022ppo}%
  \BibitemOpen
  \bibfield  {author} {\bibinfo {author} {\bibfnamefont {W.~E.}\ \bibnamefont {East}},\ }\href {\doibase 10.1103/PhysRevLett.129.141103} {\bibfield  {journal} {\bibinfo  {journal} {Phys. Rev. Lett.}\ }\textbf {\bibinfo {volume} {129}},\ \bibinfo {pages} {141103} (\bibinfo {year} {2022})},\ \Eprint {http://arxiv.org/abs/2205.03417} {arXiv:2205.03417 [hep-ph]} \BibitemShut {NoStop}%
\bibitem [{\citenamefont {Baryakhtar}\ \emph {et~al.}(2021)\citenamefont {Baryakhtar}, \citenamefont {Galanis}, \citenamefont {Lasenby},\ and\ \citenamefont {Simon}}]{Baryakhtar:2020gao}%
  \BibitemOpen
  \bibfield  {author} {\bibinfo {author} {\bibfnamefont {M.}~\bibnamefont {Baryakhtar}}, \bibinfo {author} {\bibfnamefont {M.}~\bibnamefont {Galanis}}, \bibinfo {author} {\bibfnamefont {R.}~\bibnamefont {Lasenby}}, \ and\ \bibinfo {author} {\bibfnamefont {O.}~\bibnamefont {Simon}},\ }\href {\doibase 10.1103/PhysRevD.103.095019} {\bibfield  {journal} {\bibinfo  {journal} {Phys. Rev. D}\ }\textbf {\bibinfo {volume} {103}},\ \bibinfo {pages} {095019} (\bibinfo {year} {2021})},\ \Eprint {http://arxiv.org/abs/2011.11646} {arXiv:2011.11646 [hep-ph]} \BibitemShut {NoStop}%
\bibitem [{\citenamefont {Boskovic}\ \emph {et~al.}(2019)\citenamefont {Boskovic}, \citenamefont {Brito}, \citenamefont {Cardoso}, \citenamefont {Ikeda},\ and\ \citenamefont {Witek}}]{Boskovic:2018lkj}%
  \BibitemOpen
  \bibfield  {author} {\bibinfo {author} {\bibfnamefont {M.}~\bibnamefont {Boskovic}}, \bibinfo {author} {\bibfnamefont {R.}~\bibnamefont {Brito}}, \bibinfo {author} {\bibfnamefont {V.}~\bibnamefont {Cardoso}}, \bibinfo {author} {\bibfnamefont {T.}~\bibnamefont {Ikeda}}, \ and\ \bibinfo {author} {\bibfnamefont {H.}~\bibnamefont {Witek}},\ }\href {\doibase 10.1103/PhysRevD.99.035006} {\bibfield  {journal} {\bibinfo  {journal} {Phys. Rev. D}\ }\textbf {\bibinfo {volume} {99}},\ \bibinfo {pages} {035006} (\bibinfo {year} {2019})},\ \Eprint {http://arxiv.org/abs/1811.04945} {arXiv:1811.04945 [gr-qc]} \BibitemShut {NoStop}%
\bibitem [{\citenamefont {Brito}\ \emph {et~al.}(2015)\citenamefont {Brito}, \citenamefont {Cardoso},\ and\ \citenamefont {Pani}}]{Brito:2015oca}%
  \BibitemOpen
  \bibfield  {author} {\bibinfo {author} {\bibfnamefont {R.}~\bibnamefont {Brito}}, \bibinfo {author} {\bibfnamefont {V.}~\bibnamefont {Cardoso}}, \ and\ \bibinfo {author} {\bibfnamefont {P.}~\bibnamefont {Pani}},\ }\href {\doibase 10.1007/978-3-319-19000-6} {\bibfield  {journal} {\bibinfo  {journal} {Lect. Notes Phys.}\ }\textbf {\bibinfo {volume} {906}},\ \bibinfo {pages} {pp.1} (\bibinfo {year} {2015})},\ \Eprint {http://arxiv.org/abs/1501.06570} {arXiv:1501.06570 [gr-qc]} \BibitemShut {NoStop}%
\bibitem [{\citenamefont {Linde}(1992)}]{Linde:1991sk}%
  \BibitemOpen
  \bibfield  {author} {\bibinfo {author} {\bibfnamefont {A.~D.}\ \bibnamefont {Linde}},\ }\href {\doibase 10.1016/0550-3213(92)90326-7} {\bibfield  {journal} {\bibinfo  {journal} {Nucl. Phys. B}\ }\textbf {\bibinfo {volume} {372}},\ \bibinfo {pages} {421} (\bibinfo {year} {1992})},\ \Eprint {http://arxiv.org/abs/hep-th/9110037} {arXiv:hep-th/9110037} \BibitemShut {NoStop}%
\bibitem [{\citenamefont {Brown}\ and\ \citenamefont {Dahlen}(2011)}]{Brown:2011ry}%
  \BibitemOpen
  \bibfield  {author} {\bibinfo {author} {\bibfnamefont {A.~R.}\ \bibnamefont {Brown}}\ and\ \bibinfo {author} {\bibfnamefont {A.}~\bibnamefont {Dahlen}},\ }\href {\doibase 10.1103/PhysRevLett.107.171301} {\bibfield  {journal} {\bibinfo  {journal} {Phys. Rev. Lett.}\ }\textbf {\bibinfo {volume} {107}},\ \bibinfo {pages} {171301} (\bibinfo {year} {2011})},\ \Eprint {http://arxiv.org/abs/1108.0119} {arXiv:1108.0119 [hep-th]} \BibitemShut {NoStop}%
\bibitem [{\citenamefont {Braden}\ \emph {et~al.}(2019)\citenamefont {Braden}, \citenamefont {Johnson}, \citenamefont {Peiris}, \citenamefont {Pontzen},\ and\ \citenamefont {Weinfurtner}}]{Braden:2018tky}%
  \BibitemOpen
  \bibfield  {author} {\bibinfo {author} {\bibfnamefont {J.}~\bibnamefont {Braden}}, \bibinfo {author} {\bibfnamefont {M.~C.}\ \bibnamefont {Johnson}}, \bibinfo {author} {\bibfnamefont {H.~V.}\ \bibnamefont {Peiris}}, \bibinfo {author} {\bibfnamefont {A.}~\bibnamefont {Pontzen}}, \ and\ \bibinfo {author} {\bibfnamefont {S.}~\bibnamefont {Weinfurtner}},\ }\href {\doibase 10.1103/PhysRevLett.123.031601} {\bibfield  {journal} {\bibinfo  {journal} {Phys. Rev. Lett.}\ }\textbf {\bibinfo {volume} {123}},\ \bibinfo {pages} {031601} (\bibinfo {year} {2019})},\ \bibinfo {note} {[Erratum: Phys.Rev.Lett. 129, 059901 (2022)]},\ \Eprint {http://arxiv.org/abs/1806.06069} {arXiv:1806.06069 [hep-th]} \BibitemShut {NoStop}%
\bibitem [{\citenamefont {Blanco-Pillado}\ \emph {et~al.}(2019)\citenamefont {Blanco-Pillado}, \citenamefont {Deng},\ and\ \citenamefont {Vilenkin}}]{Blanco-Pillado:2019xny}%
  \BibitemOpen
  \bibfield  {author} {\bibinfo {author} {\bibfnamefont {J.~J.}\ \bibnamefont {Blanco-Pillado}}, \bibinfo {author} {\bibfnamefont {H.}~\bibnamefont {Deng}}, \ and\ \bibinfo {author} {\bibfnamefont {A.}~\bibnamefont {Vilenkin}},\ }\href {\doibase 10.1088/1475-7516/2019/12/001} {\bibfield  {journal} {\bibinfo  {journal} {JCAP}\ }\textbf {\bibinfo {volume} {12}},\ \bibinfo {pages} {001} (\bibinfo {year} {2019})},\ \Eprint {http://arxiv.org/abs/1906.09657} {arXiv:1906.09657 [hep-th]} \BibitemShut {NoStop}%
\bibitem [{\citenamefont {Hertzberg}\ and\ \citenamefont {Yamada}(2019)}]{Hertzberg:2019wgx}%
  \BibitemOpen
  \bibfield  {author} {\bibinfo {author} {\bibfnamefont {M.~P.}\ \bibnamefont {Hertzberg}}\ and\ \bibinfo {author} {\bibfnamefont {M.}~\bibnamefont {Yamada}},\ }\href {\doibase 10.1103/PhysRevD.100.016011} {\bibfield  {journal} {\bibinfo  {journal} {Phys. Rev. D}\ }\textbf {\bibinfo {volume} {100}},\ \bibinfo {pages} {016011} (\bibinfo {year} {2019})},\ \Eprint {http://arxiv.org/abs/1904.08565} {arXiv:1904.08565 [hep-th]} \BibitemShut {NoStop}%
\bibitem [{\citenamefont {Hertzberg}\ \emph {et~al.}(2020)\citenamefont {Hertzberg}, \citenamefont {Rompineve},\ and\ \citenamefont {Shah}}]{Hertzberg:2020tqa}%
  \BibitemOpen
  \bibfield  {author} {\bibinfo {author} {\bibfnamefont {M.~P.}\ \bibnamefont {Hertzberg}}, \bibinfo {author} {\bibfnamefont {F.}~\bibnamefont {Rompineve}}, \ and\ \bibinfo {author} {\bibfnamefont {N.}~\bibnamefont {Shah}},\ }\href {\doibase 10.1103/PhysRevD.102.076003} {\bibfield  {journal} {\bibinfo  {journal} {Phys. Rev. D}\ }\textbf {\bibinfo {volume} {102}},\ \bibinfo {pages} {076003} (\bibinfo {year} {2020})},\ \Eprint {http://arxiv.org/abs/2009.00017} {arXiv:2009.00017 [hep-th]} \BibitemShut {NoStop}%
\bibitem [{\citenamefont {Braden}\ \emph {et~al.}(2023)\citenamefont {Braden}, \citenamefont {Johnson}, \citenamefont {Peiris}, \citenamefont {Pontzen},\ and\ \citenamefont {Weinfurtner}}]{Braden:2022odm}%
  \BibitemOpen
  \bibfield  {author} {\bibinfo {author} {\bibfnamefont {J.}~\bibnamefont {Braden}}, \bibinfo {author} {\bibfnamefont {M.~C.}\ \bibnamefont {Johnson}}, \bibinfo {author} {\bibfnamefont {H.~V.}\ \bibnamefont {Peiris}}, \bibinfo {author} {\bibfnamefont {A.}~\bibnamefont {Pontzen}}, \ and\ \bibinfo {author} {\bibfnamefont {S.}~\bibnamefont {Weinfurtner}},\ }\href {\doibase 10.1103/PhysRevD.107.083509} {\bibfield  {journal} {\bibinfo  {journal} {Phys. Rev. D}\ }\textbf {\bibinfo {volume} {107}},\ \bibinfo {pages} {083509} (\bibinfo {year} {2023})},\ \Eprint {http://arxiv.org/abs/2204.11867} {arXiv:2204.11867 [hep-th]} \BibitemShut {NoStop}%
\bibitem [{\citenamefont {Jenkins}\ \emph {et~al.}(2023)\citenamefont {Jenkins}, \citenamefont {Moss}, \citenamefont {Billam}, \citenamefont {Hadzibabic}, \citenamefont {Peiris},\ and\ \citenamefont {Pontzen}}]{Jenkins:2023npg}%
  \BibitemOpen
  \bibfield  {author} {\bibinfo {author} {\bibfnamefont {A.~C.}\ \bibnamefont {Jenkins}}, \bibinfo {author} {\bibfnamefont {I.~G.}\ \bibnamefont {Moss}}, \bibinfo {author} {\bibfnamefont {T.~P.}\ \bibnamefont {Billam}}, \bibinfo {author} {\bibfnamefont {Z.}~\bibnamefont {Hadzibabic}}, \bibinfo {author} {\bibfnamefont {H.~V.}\ \bibnamefont {Peiris}}, \ and\ \bibinfo {author} {\bibfnamefont {A.}~\bibnamefont {Pontzen}},\ }\href@noop {} {\  (\bibinfo {year} {2023})},\ \Eprint {http://arxiv.org/abs/2311.02156} {arXiv:2311.02156 [cond-mat.quant-gas]} \BibitemShut {NoStop}%
\bibitem [{\citenamefont {Jenkins}\ \emph {et~al.}(2024)\citenamefont {Jenkins}, \citenamefont {Braden}, \citenamefont {Peiris}, \citenamefont {Pontzen}, \citenamefont {Johnson},\ and\ \citenamefont {Weinfurtner}}]{Jenkins:2023eez}%
  \BibitemOpen
  \bibfield  {author} {\bibinfo {author} {\bibfnamefont {A.~C.}\ \bibnamefont {Jenkins}}, \bibinfo {author} {\bibfnamefont {J.}~\bibnamefont {Braden}}, \bibinfo {author} {\bibfnamefont {H.~V.}\ \bibnamefont {Peiris}}, \bibinfo {author} {\bibfnamefont {A.}~\bibnamefont {Pontzen}}, \bibinfo {author} {\bibfnamefont {M.~C.}\ \bibnamefont {Johnson}}, \ and\ \bibinfo {author} {\bibfnamefont {S.}~\bibnamefont {Weinfurtner}},\ }\href {\doibase 10.1103/PhysRevD.109.023506} {\bibfield  {journal} {\bibinfo  {journal} {Phys. Rev. D}\ }\textbf {\bibinfo {volume} {109}},\ \bibinfo {pages} {023506} (\bibinfo {year} {2024})},\ \Eprint {http://arxiv.org/abs/2307.02549} {arXiv:2307.02549 [cond-mat.quant-gas]} \BibitemShut {NoStop}%
\bibitem [{\citenamefont {Sch\"afer}\ \emph {et~al.}(2001)\citenamefont {Sch\"afer}, \citenamefont {Son}, \citenamefont {Stephanov}, \citenamefont {Toublan},\ and\ \citenamefont {Verbaarschot}}]{Schafer:2001bq}%
  \BibitemOpen
  \bibfield  {author} {\bibinfo {author} {\bibfnamefont {T.}~\bibnamefont {Sch\"afer}}, \bibinfo {author} {\bibfnamefont {D.~T.}\ \bibnamefont {Son}}, \bibinfo {author} {\bibfnamefont {M.~A.}\ \bibnamefont {Stephanov}}, \bibinfo {author} {\bibfnamefont {D.}~\bibnamefont {Toublan}}, \ and\ \bibinfo {author} {\bibfnamefont {J.~J.~M.}\ \bibnamefont {Verbaarschot}},\ }\href {\doibase 10.1016/S0370-2693(01)01265-5} {\bibfield  {journal} {\bibinfo  {journal} {Phys. Lett. B}\ }\textbf {\bibinfo {volume} {522}},\ \bibinfo {pages} {67} (\bibinfo {year} {2001})},\ \Eprint {http://arxiv.org/abs/hep-ph/0108210} {arXiv:hep-ph/0108210} \BibitemShut {NoStop}%
\bibitem [{\citenamefont {Dolan}\ and\ \citenamefont {Jackiw}(1974)}]{Dolan:1973qd}%
  \BibitemOpen
  \bibfield  {author} {\bibinfo {author} {\bibfnamefont {L.}~\bibnamefont {Dolan}}\ and\ \bibinfo {author} {\bibfnamefont {R.}~\bibnamefont {Jackiw}},\ }\href {\doibase 10.1103/PhysRevD.9.3320} {\bibfield  {journal} {\bibinfo  {journal} {Phys. Rev. D}\ }\textbf {\bibinfo {volume} {9}},\ \bibinfo {pages} {3320} (\bibinfo {year} {1974})}\BibitemShut {NoStop}%
\bibitem [{\citenamefont {Visser}(2007)}]{Visser:2007fj}%
  \BibitemOpen
  \bibfield  {author} {\bibinfo {author} {\bibfnamefont {M.}~\bibnamefont {Visser}},\ }in\ \href@noop {} {\emph {\bibinfo {booktitle} {{Kerr Fest: Black Holes in Astrophysics, General Relativity and Quantum Gravity}}}}\ (\bibinfo {year} {2007})\ \Eprint {http://arxiv.org/abs/0706.0622} {arXiv:0706.0622 [gr-qc]} \BibitemShut {NoStop}%
\bibitem [{\citenamefont {Clough}\ \emph {et~al.}(2019)\citenamefont {Clough}, \citenamefont {Ferreira},\ and\ \citenamefont {Lagos}}]{Clough:2019jpm}%
  \BibitemOpen
  \bibfield  {author} {\bibinfo {author} {\bibfnamefont {K.}~\bibnamefont {Clough}}, \bibinfo {author} {\bibfnamefont {P.~G.}\ \bibnamefont {Ferreira}}, \ and\ \bibinfo {author} {\bibfnamefont {M.}~\bibnamefont {Lagos}},\ }\href {\doibase 10.1103/PhysRevD.100.063014} {\bibfield  {journal} {\bibinfo  {journal} {Phys. Rev. D}\ }\textbf {\bibinfo {volume} {100}},\ \bibinfo {pages} {063014} (\bibinfo {year} {2019})},\ \Eprint {http://arxiv.org/abs/1904.12783} {arXiv:1904.12783 [gr-qc]} \BibitemShut {NoStop}%
\bibitem [{\citenamefont {Bamber}\ \emph {et~al.}(2021)\citenamefont {Bamber}, \citenamefont {Clough}, \citenamefont {Ferreira}, \citenamefont {Hui},\ and\ \citenamefont {Lagos}}]{Bamber:2020bpu}%
  \BibitemOpen
  \bibfield  {author} {\bibinfo {author} {\bibfnamefont {J.}~\bibnamefont {Bamber}}, \bibinfo {author} {\bibfnamefont {K.}~\bibnamefont {Clough}}, \bibinfo {author} {\bibfnamefont {P.~G.}\ \bibnamefont {Ferreira}}, \bibinfo {author} {\bibfnamefont {L.}~\bibnamefont {Hui}}, \ and\ \bibinfo {author} {\bibfnamefont {M.}~\bibnamefont {Lagos}},\ }\href {\doibase 10.1103/PhysRevD.103.044059} {\bibfield  {journal} {\bibinfo  {journal} {Phys. Rev. D}\ }\textbf {\bibinfo {volume} {103}},\ \bibinfo {pages} {044059} (\bibinfo {year} {2021})},\ \Eprint {http://arxiv.org/abs/2011.07870} {arXiv:2011.07870 [gr-qc]} \BibitemShut {NoStop}%
\bibitem [{\citenamefont {Hui}\ \emph {et~al.}(2019)\citenamefont {Hui}, \citenamefont {Kabat}, \citenamefont {Li}, \citenamefont {Santoni},\ and\ \citenamefont {Wong}}]{Hui:2019aqm}%
  \BibitemOpen
  \bibfield  {author} {\bibinfo {author} {\bibfnamefont {L.}~\bibnamefont {Hui}}, \bibinfo {author} {\bibfnamefont {D.}~\bibnamefont {Kabat}}, \bibinfo {author} {\bibfnamefont {X.}~\bibnamefont {Li}}, \bibinfo {author} {\bibfnamefont {L.}~\bibnamefont {Santoni}}, \ and\ \bibinfo {author} {\bibfnamefont {S.~S.~C.}\ \bibnamefont {Wong}},\ }\href {\doibase 10.1088/1475-7516/2019/06/038} {\bibfield  {journal} {\bibinfo  {journal} {JCAP}\ }\textbf {\bibinfo {volume} {06}},\ \bibinfo {pages} {038} (\bibinfo {year} {2019})},\ \Eprint {http://arxiv.org/abs/1904.12803} {arXiv:1904.12803 [gr-qc]} \BibitemShut {NoStop}%
\bibitem [{\citenamefont {Vieira}\ \emph {et~al.}(2014)\citenamefont {Vieira}, \citenamefont {Bezerra},\ and\ \citenamefont {Muniz}}]{Vieira:2014waa}%
  \BibitemOpen
  \bibfield  {author} {\bibinfo {author} {\bibfnamefont {H.~S.}\ \bibnamefont {Vieira}}, \bibinfo {author} {\bibfnamefont {V.~B.}\ \bibnamefont {Bezerra}}, \ and\ \bibinfo {author} {\bibfnamefont {C.~R.}\ \bibnamefont {Muniz}},\ }\href {\doibase 10.1016/j.aop.2014.07.011} {\bibfield  {journal} {\bibinfo  {journal} {Annals Phys.}\ }\textbf {\bibinfo {volume} {350}},\ \bibinfo {pages} {14} (\bibinfo {year} {2014})},\ \Eprint {http://arxiv.org/abs/1401.5397} {arXiv:1401.5397 [gr-qc]} \BibitemShut {NoStop}%
\bibitem [{\citenamefont {Hortacsu}(2012)}]{Hortacsu:2011rr}%
  \BibitemOpen
  \bibfield  {author} {\bibinfo {author} {\bibfnamefont {M.}~\bibnamefont {Hortacsu}},\ }\href {\doibase 10.1142/9789814417532_0002} {\ ,\ \bibinfo {pages} {23} (\bibinfo {year} {2012})},\ \Eprint {http://arxiv.org/abs/1101.0471} {arXiv:1101.0471 [math-ph]} \BibitemShut {NoStop}%
\bibitem [{\citenamefont {Aurrekoetxea}\ \emph {et~al.}(2023{\natexlab{a}})\citenamefont {Aurrekoetxea}, \citenamefont {Bamber}, \citenamefont {Brady}, \citenamefont {Clough}, \citenamefont {Helfer}, \citenamefont {Marsden}, \citenamefont {Traykova},\ and\ \citenamefont {Wang}}]{Aurrekoetxea:2023fhl}%
  \BibitemOpen
  \bibfield  {author} {\bibinfo {author} {\bibfnamefont {J.~C.}\ \bibnamefont {Aurrekoetxea}}, \bibinfo {author} {\bibfnamefont {J.}~\bibnamefont {Bamber}}, \bibinfo {author} {\bibfnamefont {S.~E.}\ \bibnamefont {Brady}}, \bibinfo {author} {\bibfnamefont {K.}~\bibnamefont {Clough}}, \bibinfo {author} {\bibfnamefont {T.}~\bibnamefont {Helfer}}, \bibinfo {author} {\bibfnamefont {J.}~\bibnamefont {Marsden}}, \bibinfo {author} {\bibfnamefont {D.}~\bibnamefont {Traykova}}, \ and\ \bibinfo {author} {\bibfnamefont {Z.}~\bibnamefont {Wang}},\ }\href@noop {} {\  (\bibinfo {year} {2023}{\natexlab{a}})},\ \Eprint {http://arxiv.org/abs/2308.08299} {arXiv:2308.08299 [gr-qc]} \BibitemShut {NoStop}%
\bibitem [{\citenamefont {Andrade}\ \emph {et~al.}(2021)\citenamefont {Andrade} \emph {et~al.}}]{Andrade:2021rbd}%
  \BibitemOpen
  \bibfield  {author} {\bibinfo {author} {\bibfnamefont {T.}~\bibnamefont {Andrade}} \emph {et~al.},\ }\href {\doibase 10.21105/joss.03703} {\bibfield  {journal} {\bibinfo  {journal} {J. Open Source Softw.}\ }\textbf {\bibinfo {volume} {6}},\ \bibinfo {pages} {3703} (\bibinfo {year} {2021})},\ \Eprint {http://arxiv.org/abs/2201.03458} {arXiv:2201.03458 [gr-qc]} \BibitemShut {NoStop}%
\bibitem [{mov(2023)}]{movie}%
  \BibitemOpen
  \href@noop {} {\enquote {\bibinfo {title} {Symmetry restoration and vacuum decay from accretion around black holes},}\ } (\bibinfo {year} {2023}),\ \bibinfo {note} {\url{https://youtu.be/c2vcNoX8fVs}}\BibitemShut {NoStop}%
\bibitem [{\citenamefont {Barranco}\ \emph {et~al.}(2012)\citenamefont {Barranco}, \citenamefont {Bernal}, \citenamefont {Degollado}, \citenamefont {Diez-Tejedor}, \citenamefont {Megevand}, \citenamefont {Alcubierre}, \citenamefont {Nunez},\ and\ \citenamefont {Sarbach}}]{Barranco:2012qs}%
  \BibitemOpen
  \bibfield  {author} {\bibinfo {author} {\bibfnamefont {J.}~\bibnamefont {Barranco}}, \bibinfo {author} {\bibfnamefont {A.}~\bibnamefont {Bernal}}, \bibinfo {author} {\bibfnamefont {J.~C.}\ \bibnamefont {Degollado}}, \bibinfo {author} {\bibfnamefont {A.}~\bibnamefont {Diez-Tejedor}}, \bibinfo {author} {\bibfnamefont {M.}~\bibnamefont {Megevand}}, \bibinfo {author} {\bibfnamefont {M.}~\bibnamefont {Alcubierre}}, \bibinfo {author} {\bibfnamefont {D.}~\bibnamefont {Nunez}}, \ and\ \bibinfo {author} {\bibfnamefont {O.}~\bibnamefont {Sarbach}},\ }\href {\doibase 10.1103/PhysRevLett.109.081102} {\bibfield  {journal} {\bibinfo  {journal} {Phys. Rev. Lett.}\ }\textbf {\bibinfo {volume} {109}},\ \bibinfo {pages} {081102} (\bibinfo {year} {2012})},\ \Eprint {http://arxiv.org/abs/1207.2153} {arXiv:1207.2153 [gr-qc]} \BibitemShut {NoStop}%
\bibitem [{\citenamefont {Cespedes}\ \emph {et~al.}(2021)\citenamefont {Cespedes}, \citenamefont {de~Alwis}, \citenamefont {Muia},\ and\ \citenamefont {Quevedo}}]{Cespedes:2020xpn}%
  \BibitemOpen
  \bibfield  {author} {\bibinfo {author} {\bibfnamefont {S.}~\bibnamefont {Cespedes}}, \bibinfo {author} {\bibfnamefont {S.~P.}\ \bibnamefont {de~Alwis}}, \bibinfo {author} {\bibfnamefont {F.}~\bibnamefont {Muia}}, \ and\ \bibinfo {author} {\bibfnamefont {F.}~\bibnamefont {Quevedo}},\ }\href {\doibase 10.1103/PhysRevD.104.026013} {\bibfield  {journal} {\bibinfo  {journal} {Phys. Rev. D}\ }\textbf {\bibinfo {volume} {104}},\ \bibinfo {pages} {026013} (\bibinfo {year} {2021})},\ \Eprint {http://arxiv.org/abs/2011.13936} {arXiv:2011.13936 [hep-th]} \BibitemShut {NoStop}%
\bibitem [{\citenamefont {Bamber}\ \emph {et~al.}(2023)\citenamefont {Bamber}, \citenamefont {Aurrekoetxea}, \citenamefont {Clough},\ and\ \citenamefont {Ferreira}}]{Bamber:2022pbs}%
  \BibitemOpen
  \bibfield  {author} {\bibinfo {author} {\bibfnamefont {J.}~\bibnamefont {Bamber}}, \bibinfo {author} {\bibfnamefont {J.~C.}\ \bibnamefont {Aurrekoetxea}}, \bibinfo {author} {\bibfnamefont {K.}~\bibnamefont {Clough}}, \ and\ \bibinfo {author} {\bibfnamefont {P.~G.}\ \bibnamefont {Ferreira}},\ }\href {\doibase 10.1103/PhysRevD.107.024035} {\bibfield  {journal} {\bibinfo  {journal} {Phys. Rev. D}\ }\textbf {\bibinfo {volume} {107}},\ \bibinfo {pages} {024035} (\bibinfo {year} {2023})},\ \Eprint {http://arxiv.org/abs/2210.09254} {arXiv:2210.09254 [gr-qc]} \BibitemShut {NoStop}%
\bibitem [{\citenamefont {Aurrekoetxea}\ \emph {et~al.}(2023{\natexlab{b}})\citenamefont {Aurrekoetxea}, \citenamefont {Clough}, \citenamefont {Bamber},\ and\ \citenamefont {Ferreira}}]{Aurrekoetxea:2023jwk}%
  \BibitemOpen
  \bibfield  {author} {\bibinfo {author} {\bibfnamefont {J.~C.}\ \bibnamefont {Aurrekoetxea}}, \bibinfo {author} {\bibfnamefont {K.}~\bibnamefont {Clough}}, \bibinfo {author} {\bibfnamefont {J.}~\bibnamefont {Bamber}}, \ and\ \bibinfo {author} {\bibfnamefont {P.~G.}\ \bibnamefont {Ferreira}},\ }\href@noop {} {\  (\bibinfo {year} {2023}{\natexlab{b}})},\ \Eprint {http://arxiv.org/abs/2311.18156} {arXiv:2311.18156 [gr-qc]} \BibitemShut {NoStop}%
\end{thebibliography}%

\end{document}